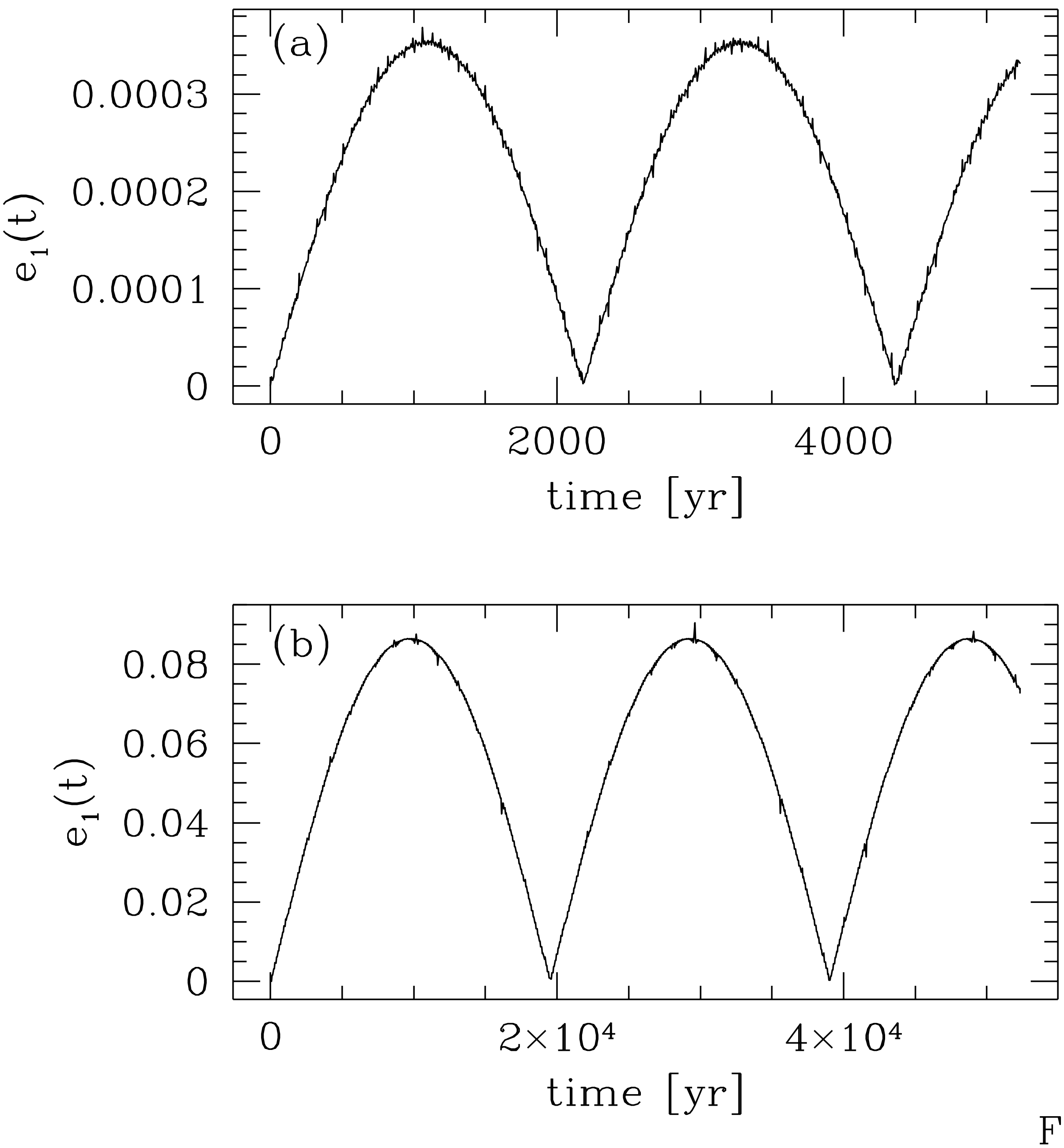

FIG. 1

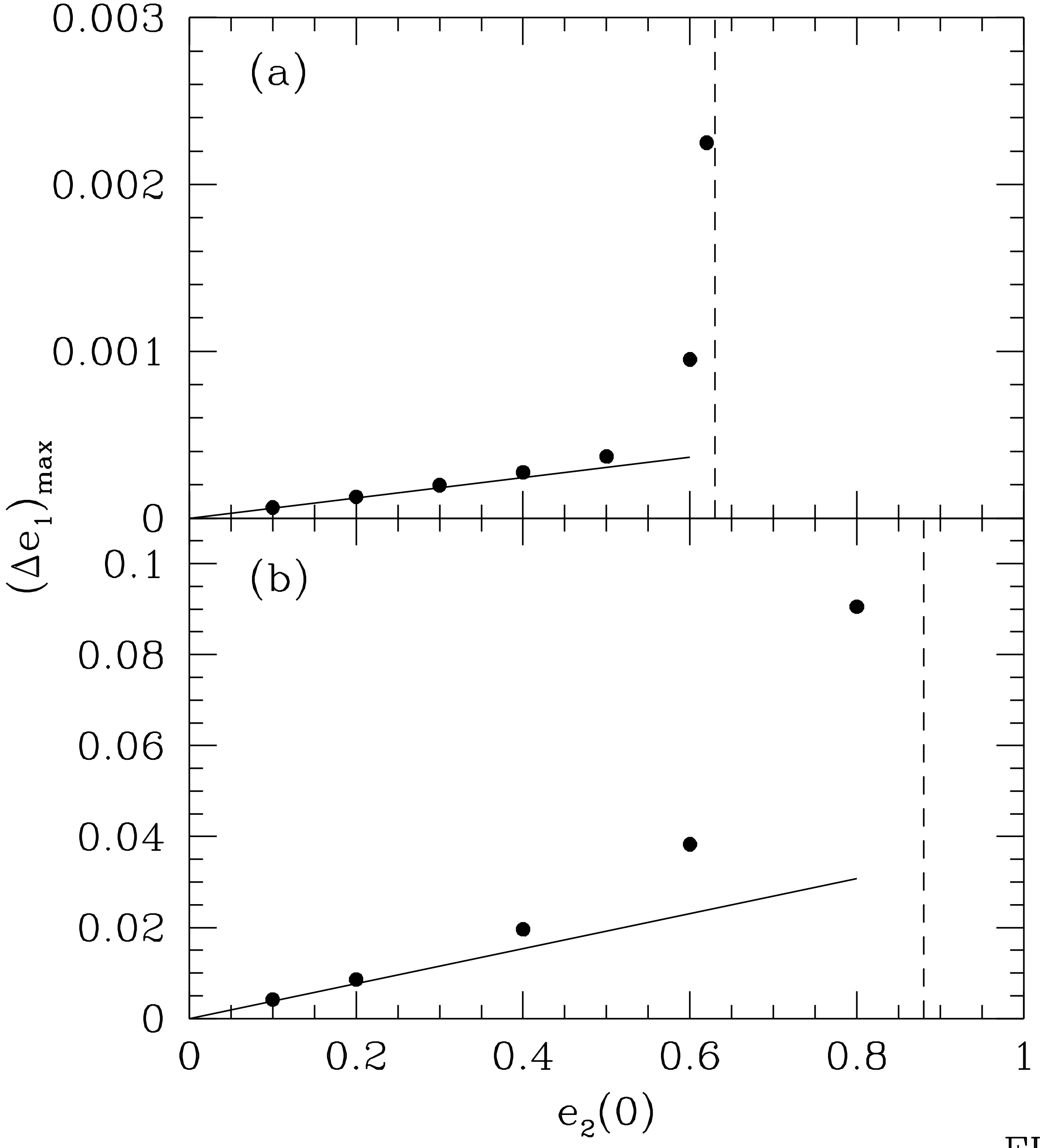

FIG. 2

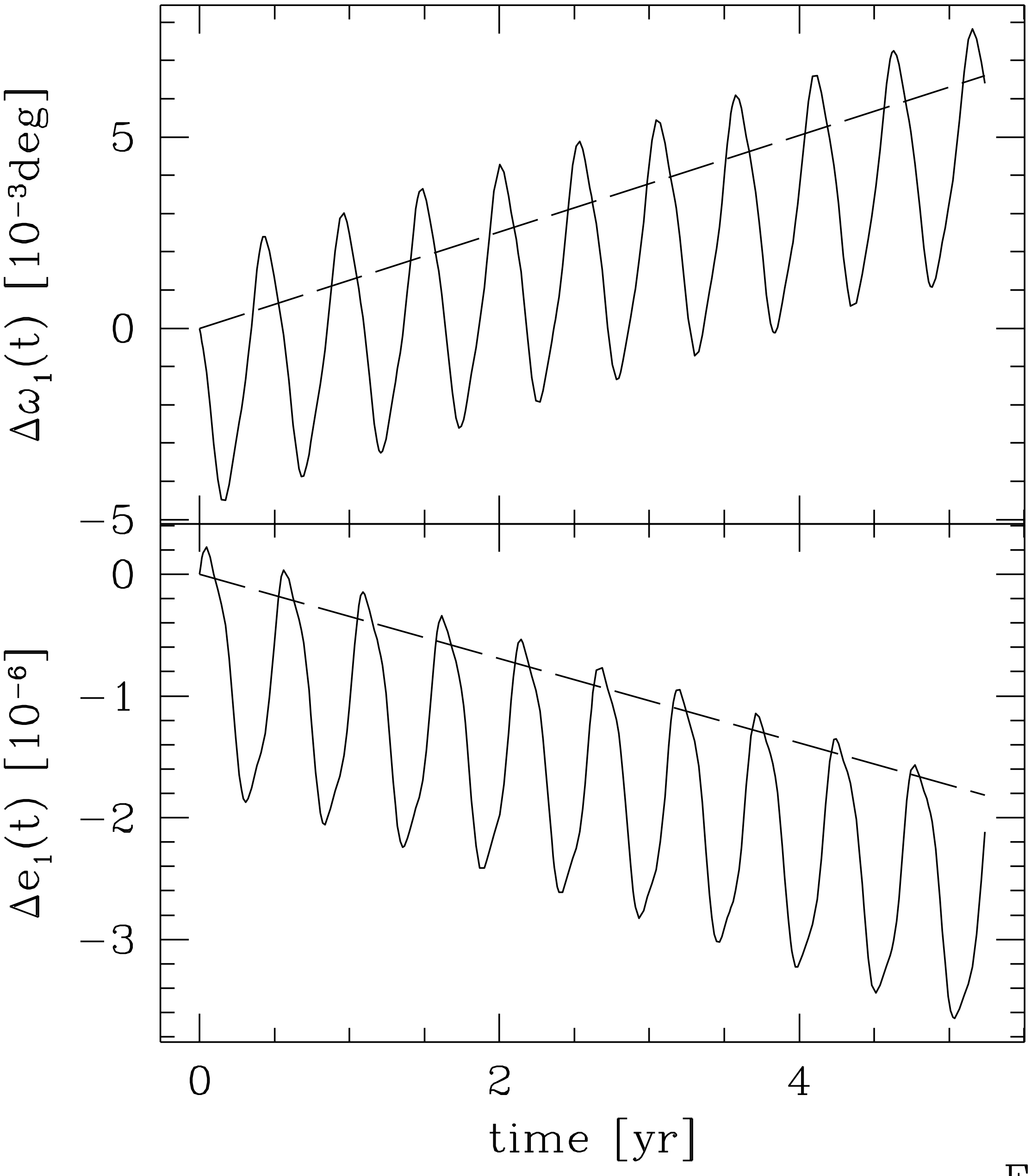

FIG. 3

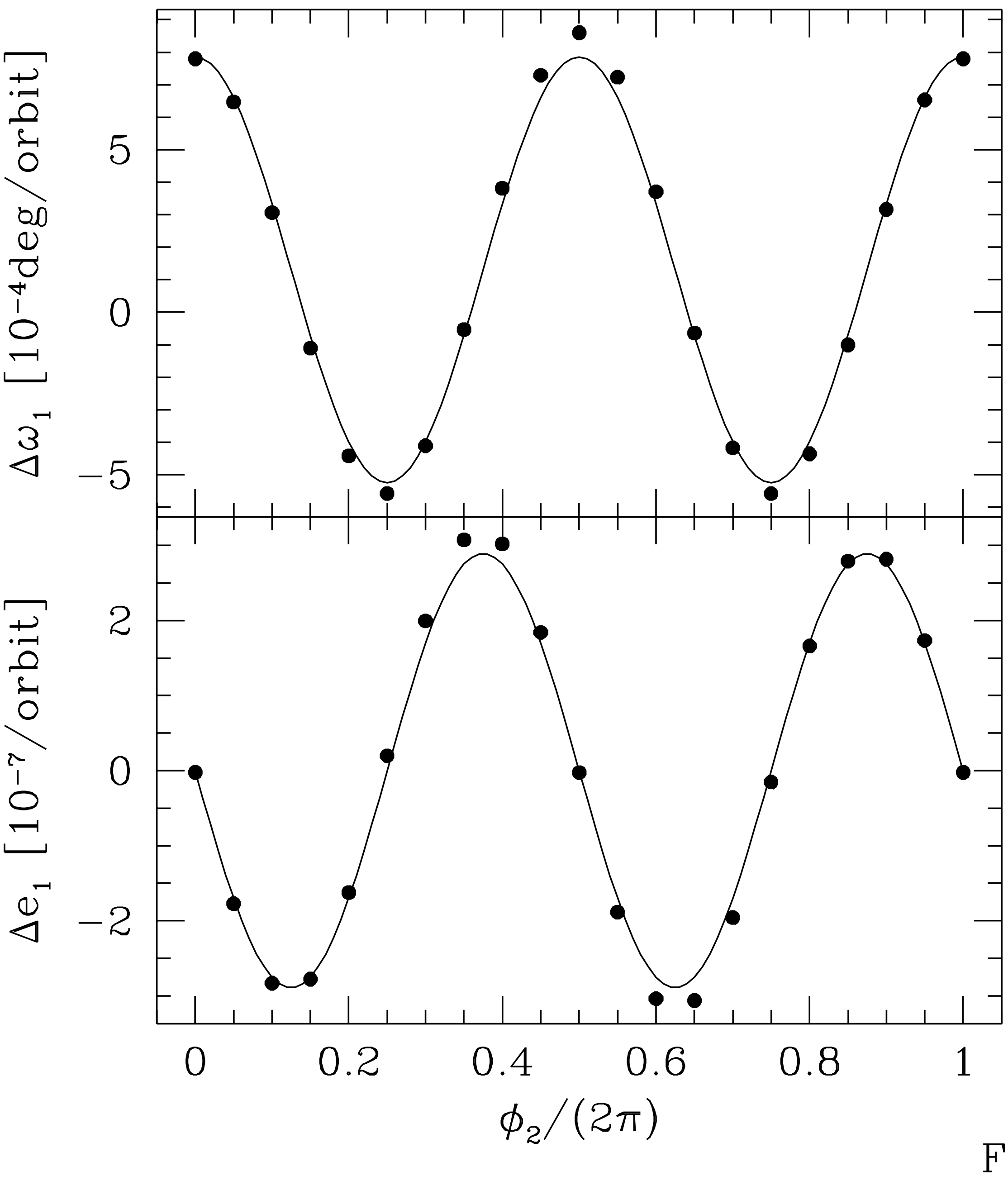

FIG. 4

# Is There a Planet in the PSR 1620-26 Triple System?

Frederic A. Rasio[1]
Institute for Advanced Study, Olden Lane, Princeton, NJ 08540

## ABSTRACT

The unusually large eccentricity ($e_1 = 0.025$) of the low-mass binary millisecond pulsar PSR B1620-26 can be explained naturally as arising from the secular perturbation of a second, more distant companion. Such a triple configuration has been proposed recently as the most likely cause of the anomalous second period derivative of the pulsar. The current timing data are consistent with a second companion mass $m_2$ as low as $\sim 10^{-3}\,M_\odot$, i.e., comparable to that of Jupiter. However, *if* the eccentricity is indeed produced by secular perturbations, then the second companion must be another star, most likely of mass $m_2 \lesssim 1 M_\odot$ and in a very eccentric ($e_2 \gtrsim 0.5$) orbit of period $P_2 \sim 10^2$–$10^3\,$yr. A second companion of planetary mass cannot induce the observed eccentricity. Independent of the mass of the second companion, small changes in the binary pulsar's orbit should become detectable with just a few additional years of timing data. This detection would provide direct confirmation of the triple nature of the system, and an accurate measurement of the effects would place important new constraints on the orbital parameters.

## 1. Introduction

The millisecond pulsar PSR B1620-26 in the globular cluster M4 has a low-mass companion (mass $m_1 \approx 0.3\,M_\odot$ for a pulsar mass $m_p = 1.35\,M_\odot$) in a nearly circular orbit of period $P_1 = 0.524\,$yr (Lyne et al. 1988; McKenna & Lyne 1988). The eccentricity $e_1 = 0.0253$, although small, is several orders of magnitude larger than observed in most other low-mass binary millisecond pulsars (Thorsett, Arzoumanian, & Taylor 1993, hereafter TAT). In addition, the timing data indicate the presence of a very large second derivative of the pulse period, $\ddot{P} = -2.3 \times 10^{-27}\,\mathrm{s}^{-1}$ (Backer 1993). This is many orders of magnitude larger than would be expected from a varying acceleration caused by either the mean gravitational field of M4 or a nearby passing star. Instead, the most likely explanation is that the pulsar has a second, more distant orbital companion (Backer 1993;

[1]Hubble Fellow.

Backer, Foster, & Sallmen 1993; TAT). Such hierarchical triple systems are expected to be produced quite easily in globular clusters through dynamical interactions between binaries (Mikkola 1984; Hut 1992).

The present timing data are consistent with a second companion mass anywhere in the range $10^{-3}\,M_\odot \lesssim m_2 \lesssim 1\,M_\odot$ with a corresponding orbital period $10\,\mathrm{yr} \lesssim P_2 \lesssim 10^3\,\mathrm{yr}$. This opens the exciting possibility that the second companion could be a Jupiter-like planet (Backer 1993; Sigurdsson 1993; TAT), although it could also be another star (main sequence, white dwarf, or even another neutron star). Planets have been detected in orbit around at least one other millisecond pulsar, PSR B1257+12 (Wolszczan & Frail 1992).

If PSR B1620-26 is indeed in a triple system, then small perturbations of the (inner) binary pulsar's orbit should be induced by the presence of the second (more distant) companion. In §2 below, we show that the unusually large eccentricity of the binary pulsar can be explained naturally as arising from these perturbations, but only if the second companion is a star and *not* a planet. In either case, small changes in the eccentricity and longitude of pericenter of the binary pulsar should become detectable with a few additional years of timing data (§3). Such a detection would provide independent confirmation that the first pulsar in a hierarchical triple system has indeed been discovered. The situation is similar to that of PSR B1257+12, where the recent detection of orbital perturbation effects has confirmed the presence of (at least) two planets in orbit around the pulsar (Rasio et al. 1992; Wolszczan 1994).

## 2. Secular Perturbations

Assume that the binary pulsar had a very low eccentricity $e_1(0) \ll e_1(t_p) = 0.025$ at the time it acquired its second companion. Standard secular perturbation theory of orbital mechanics (e.g., Brouwer & Clemence 1961) can be used to calculate the subsequent evolution of the system over a time $t_p \gg P_2 > P_1$. The particular solution for the eccentricity $e_1(t)$ of the inner orbit corresponding to an initial condition with $e_1(0) = 0$ takes the form

$$e_1(t) = \mathcal{F}(q_1, q_2, \alpha)\, e_2(0)\, [1 - \cos(gt)]^{1/2} \tag{1}$$

Here $\mathcal{F}$ is a dimensionless function of $q_j = m_j/m_p$ and $\alpha = a_1/a_2 < 1$. The angular frequency $g = g_1 - g_2$, where $g_1$ and $g_2$ are the eigenvalues of the secular perturbation matrix. An expression for $\mathcal{F}$ can be written explicitly in terms of Laplace coefficients $b^{(j)}_{3/2}(\alpha)$ (Brouwer & Clemence 1961; a useful summary of all relevant expressions is given by Dermott & Nicholson 1986). For $\alpha \ll 1$ and $q_2/q_1 \gg \alpha^{1/2}$ we find $\mathcal{F} \sim \alpha$ (independent of $q_2$) and $2\pi/g \sim P_1 q_2^{-1}\alpha^{-3}$, whereas for $\alpha \ll 1$ and $q_2/q_1 \ll \alpha^{1/2}$, $\mathcal{F} \sim q_2\alpha^{1/2}$ and $2\pi/g \sim P_1\alpha^{-7/2}$. Other treatments of secular evolution in triple systems are given by Mazeh & Shaham (1979), and Bailyn (1987).

The analytic solution (1) is valid only for small eccentricities and small relative inclination. If the system was formed through a dynamical interaction, the eccentricity and inclination of the outer orbit are likely to be quite large. Numerical integrations of the three-body problem must then be used. We have constructed a large number of numerical solutions covering the full range of parameter values allowed by the current timing data (Rasio 1994). We find that the form of the solution remains generally similar to that of equation (1), with $e_1(t)$ returning to zero periodically every $\sim 10^3$–$10^6$ yr. However, both the period and amplitude of the solutions can deviate significantly from their analytic values when $e_2$ or the inclination is large.

Our most important result is illustrated in figures 1 and 2. When the mass of the second companion is $\lesssim 0.1\,M_\odot$, the maximum eccentricity $(\Delta e_1)_{max}$ induced in the binary pulsar's orbit remains always smaller than the value $e_1 = 0.025$ observed today. For a typical planetary case, with $m_2 = 80\,M_\oplus$ and $P_2 = 10\,$yr (cf. TAT), we get $(\Delta e_1)_{max} \lesssim 0.001$ for all $e_2$. In contrast, for a typical stellar case, with $m_2 = 0.8\,M_\odot$ and $P_2 = 120\,$yr (TAT), we find that $(\Delta e_1)_{max} > 0.025$ can be obtained easily for $e_2 > 0.4$. We conclude that, *if the eccentricity of the binary pulsar's orbit has been induced by secular perturbations, then the second companion must be a star rather than a planet*.

Note that, according to equation (1), a larger amplitude $(\Delta e_1)_{max}$ can always be produced by invoking a larger eccentricity $e_2$ for the outer orbit. However, *dynamical stability* (e.g., Black 1982) requires that $(1 - e_2)a_2/a_1 \gtrsim 3$–4, placing a direct limit on how large $e_2$ can be for a given $\alpha$ (cf. Fig. 2). For example, we note that the solution with $m_2 = 0.8\,M_\odot$, $P_2 = 120\,$yr, and $e_2 = 0.9$ mentioned by TAT is dynamically unstable.

A number of alternative explanations for the anomalous $e_1$ are possible, depending on the triple formation process. If the second companion is a star, then the most natural formation process is through an interaction between the (pre-existing) binary pulsar and a much wider primordial binary containing two main-sequence stars. The typical outcome of such an interaction is a triple system containing one of the main-sequence stars in a wide orbit around the binary pulsar. The orbital eccentricity of the binary pulsar could have been perturbed during this interaction. However, this would require a close passage (within $r \lesssim 5a_1$) by one of the main-sequence stars, implying a high probability of disrupting the binary pulsar completely. A later close encounter with a passing star in the cluster is even less likely, since it has a high probability of also ejecting the second companion.

Sigurdsson (1993) describes a formation scenario where a pre-existing neutron-star-white-dwarf binary has an interaction with a turn-off main-sequence star with a planet. The white dwarf is ejected and the main-sequence star and planet remain in orbit around the neutron star. As the main-sequence star evolves, mass transfer onto the neutron star leads to the formation of the currently observed millisecond pulsar. In this case, the orbit of the binary pulsar should have been circularized very efficiently by tidal interactions during the mass-transfer phase, leaving a residual eccentricity $e_1 \sim 10^{-4}$ (Phinney 1992). According to the above results, the planet cannot later induce an eccentricity nearly as large as observed today.

However, Sigurdsson (1993) also points out that because of interactions with passing stars, the standard secular perturbation result (eq. [1]) may not apply over a timescale comparable to the age of the system. If each interaction arbitrarily "resets" the initial conditions for the secular perturbation problem, then over a timescale $t_p \gg 2\pi/g$ the eccentricity $e_1$ may grow in a random walk manner, with $e_1(t) \sim (\Delta e_1)_{max} N_{int}^{1/2}$, where $N_{int} = t/t_{int}$ is the average number of interactions in a time $t$. The mean time between interactions for an object of size $a_2$ near the center of M4 (density $\rho = 10^4 \rho_4\, M_\odot\, \mathrm{pc}^{-3}$ and velocity dispersion $\sigma = 5\sigma_5\,\mathrm{km\,s^{-1}}$) is $t_{int} \sim 10^8\,\mathrm{yr}\,\rho_4^{-1}\sigma_5 (a_2/10\,\mathrm{au})^{-1}$. This is certainly not much shorter than the age of the system, $t_p \lesssim 10^9\,\mathrm{yr}$ (TAT). Even with optimistic values for all parameters, we get $N_{int} \lesssim 100$ which is not sufficient to produce $e_1(t_p) = 0.025$. Moreover, since the outer orbit for a planet is very "soft" (binding energy much smaller than the typical energy of a cluster star), each interaction has a significant probability of disrupting the system. Thus a number of interactions as large as 100 would leave a very low probability of survival for the triple configuration (TAT also use this likely small value of $t_{int}/t_p$ to argue against a planet).

## 3. Short-Term Perturbations

Currently available timing data for PSR B1620-26 have been obtained over a time $t_{obs} \approx 5\,\mathrm{yr}$. Clearly, we have $P_1 \ll t_{obs} \ll P_2$ for most allowed configurations. Therefore, to a good approximation, the secular perturbations observable today or over the next few years can be calculated assuming that the second companion has a *fixed position in space*. The components of the perturbing (tidal) force per unit mass acting on the pulsar are then

$$R = \frac{G m_2 r}{r_2^3}[3\sin^2\theta_2 \cos^2(\phi_2 - v) - 1] \tag{2}$$

in the radial direction ($\hat{\mathbf{r}}$),

$$B = \frac{G m_2 r}{r_2^3}\sin^2\theta_2 \cos(\phi_2 - v)\sin(\phi_2 - v) \tag{3}$$

in the perpendicular direction ($\hat{\phi}$), and

$$N = \frac{G m_2 r}{r_2^3}\, 3\sin\theta_2\cos\theta_2\cos(\phi_2 - v) \tag{4}$$

in the vertical direction ($\hat{\mathbf{z}}$). Here $r_2$, $\theta_2$ and $\phi_2$ are the fixed spherical polar coordinates of the second companion (with the origin at the center of mass of the binary pulsar and $\phi_2$ measured from pericenter in the orbital plane), and $r$ and $v$ are the pulsar's radius and true anomaly. We have assumed that $r \ll r_2$.

Using Lagrange's planetary equations (e.g., Danby 1988), one can calculate the changes in $\omega_1$ and $e_1$, integrated over one period of the inner orbit. For small $e_1$ we find

$$\Delta\omega_1 = 3\pi\,\eta \left[\sin^2\theta_2(5\cos^2\phi_2 - 1) - 1\right], \tag{5}$$

and

$$\Delta e_1 = \frac{-15\pi}{2}\,\eta\, e_1 \sin^2\theta_2 \sin(2\phi_2), \tag{6}$$

where $\eta = [m_2/(m_p + m_1)](a_1/r_2)^3$. These results are illustrated in figures 3 and 4, where we also demonstrate very good agreement with numerical solutions. As an additional check, one can easily verify that integrating equation (5) over $\phi_2$ yields the familiar result for the precession of an orbit due to a distant circular ring. For a noncoplanar system the perturbation of the inclination is

$$\Delta i_1 = \frac{3\pi}{2}\,\eta\,\sin(2\theta_2)\cos(\omega_1 + \phi_2), \tag{7}$$

which may be detectable through a change in the pulsar's projected semi-major axis $a_p \sin i$. The longitude of the node is also perturbed, but this quantity is essentially unmeasurable for a binary pulsar. There is no perturbation of the semi-major axis (or orbital period).

The measured values of the various derivatives of the pulse frequency $f$ can be used to constrain the amplitude of the changes predicted by equations (5)–(7). A complete analysis of the problem will be presented elsewhere (Rasio 1994). Here, for simplicity, consider the case of a nearly-circular outer orbit ($e_2 \approx 0$) and assume that $\sin i_1 = \sin\theta_2 = 1$. Since $\dot{f}^2/f \ll \ddot{f}$ and $|\dot{f}\ddot{f}/f| \ll |\dddot{f}|$, we can write the acceleration-induced frequency derivatives as

$$\dot{f} \;=\; f\frac{\mathbf{a}\cdot\mathbf{n}}{c} = -\frac{f}{c}\frac{Gm_2}{a_2^2}\sin(\omega_1+\phi_2) \tag{8}$$

$$\ddot{f} \;=\; f\frac{\dot{\mathbf{a}}\cdot\mathbf{n}}{c} = \frac{f}{c}\frac{G^{3/2}m_2}{a_2^{7/2}}(m_p+m_1+m_2)^{1/2}\cos(\omega_1+\phi_2) \tag{9}$$

$$\dddot{f} \;=\; f\frac{\ddot{\mathbf{a}}\cdot\mathbf{n}}{c} = \frac{f}{c}\frac{G^2m_2}{a_2^5}(m_p+m_1+m_2)\sin(\omega_1+\phi_2) \tag{10}$$

Here $\omega_1 = 117\,\mathrm{deg}$ is the longitude of pericenter measured from the ascending node ($\pi/2-\omega_1-\phi_2$ is the angle between the line of sight and the direction to the second companion), $\mathbf{a}$ is the acceleration of the binary pulsar, and $\mathbf{n}$ is a unit vector in the direction of the line of sight.

For $m_2 \ll m_p + m_1$, it is straightforward to solve the system of equations (8)–(10) for the three unknowns $m_2$, $a_2$, and $\phi_2$, given measured values of the three frequency derivatives. In general, a few iterations over $m_2$ may be needed. If we assume that the present value of $\dot{f}$ is determined predominantly by the acceleration of the pulsar (TAT), and adopt the preliminary value of $\dddot{f} = 10^{-32}\,\mathrm{s}^{-4}$ reported by Backer (1994), we find the solution $m_1 = 1.1\times10^{-2}\,M_\odot$, $a_2 = 5.2\times10^{14}\,\mathrm{cm}$, and $\phi_2 = -94\,\mathrm{deg}$. The corresponding rates of change given by equations (5) and (6) are $\dot{\omega}_1 = \Delta\omega_1/P_1 = -1.4\times10^{-4}\,\mathrm{deg\,yr^{-1}}$, $\dot{e}_1 = \Delta e_1/P_1 = -1.1\times10^{-8}\,\mathrm{yr^{-1}}$. These values should be compared to the current ($1\sigma$) upper limits derived from the timing data, $|\dot{\omega}_1| < 3\times10^{-4}\,\mathrm{deg\,yr^{-1}}$ and $|\dot{e}_1| < 10^{-6}\,\mathrm{yr^{-1}}$ (S. Thorsett, private communication). Michel (1994) has calculated more general zero-inclination solutions for $e_2 \neq 0$. In this case the longitude of pericenter $\omega_2$ of the outer orbit appears as an additional free parameter. For all solutions given by Michel (varying both $e_2$ and $\omega_2$), we find that equations (5) and (6) predict $-1.6\times10^{-4}\,\mathrm{deg\,yr^{-1}} < \dot{\omega}_1 < +0.9\times10^{-4}\,\mathrm{deg\,yr^{-1}}$ and $-0.9\times10^{-7}\,\mathrm{yr^{-1}} < \dot{e}_1 < +0.7\times10^{-7}\,\mathrm{yr^{-1}}$. The solutions have $m_2$ varying in the range $10^{-3}\,M_\odot$ to $1\,M_\odot$ and $P_2$ between 50 yr and 1300 yr. If we except special values of the angles, we find that most solutions give $-1.6\times10^{-4}\,\mathrm{deg\,yr^{-1}} < \dot{\omega}_1 < -1.1\times10^{-4}\,\mathrm{deg\,yr^{-1}}$, which is just below the present upper limit. Thus we expect $\dot{\omega}_1$, in particular, to become measurable very soon. For comparison, the general relativistic precession rate is $\dot{\omega}_{1GR} = 4.5 \times 10^{-5}\,\mathrm{deg\,yr^{-1}}[(m_p + m_1)/1.7\,M_\odot]^{2/3}$.

I am very grateful to S. Thorsett for communicating the results of observations in progress and for many useful discussions, and to P. Nicholson for a thorough reading of the manuscript. I also thank D. Backer, J. Bahcall, C. Bailyn, P. Hut, R. Malhotra, C. Michel, and S. Sigurdsson for helpful discussions and comments. The hospitality of the Aspen Center for Physics is also gratefully acknowledged. This work has been supported by a Hubble Fellowship, funded by NASA through Grant HF-1037.01-92A from the Space Telescope Science Institute, which is operated by AURA, Inc., under contract NAS5-26555.

## REFERENCES

Backer, D. C. 1993, in Planets around Pulsars, ed. J. A. Phillips et al. (ASP Conf. Ser., 36), 11

Backer, D. C. 1994, talk presented at the Aspen Winter Physics Conf. on Millisecond Pulsars: The Decade of Surprise

Backer, D. C., Foster, R. S., & Sallmen, S. 1993, Nature, 365, 817

Bailyn, C. D. 1987, ApJ, 317, 737

Black, D. C. 1982, AJ, 87, 1333

Brouwer, D., & Clemence, G. M. 1961, Methods of Celestial Mechanics (New York: Academic)

Danby, J. M. A. 1988, Fundamentals of Celestial Mechanics, 2nd edition (Richmond: Willmann-Bell)

Dermott, S. F., & Nicholson, P. D. 1986, Nature, 319, 115

Hut, P. 1992, in X-ray Binaries and Recycled Pulsars, eds. E. P. J. van den Heuvel & S. A. Rappaport (Dordrecht: Kluwer), 317

Lyne, A. G., Biggs, J. D., Brinklow, A., Ashworth, M., & McKenna, J. 1988, Nature, 332, 45

Mazeh, T., & Shaham, J. 1979, A&A, 77, 145

McKenna, J., & Lyne, A. G. 1988, Nature, 336, 226; erratum, 336, 698

Michel, F. C. 1994, preprint

Mikkola, S. 1984, MNRAS, 208, 75

Phinney, E. S. 1992, Phil. Trans. R. Soc. Lond. A, 341, 39

Rasio, F. A. 1994, in preparation

Rasio, F. A., Nicholson, P. D., Shapiro, S. L., & Teukolsky, S. A. 1992, Nature, 355, 325

Sigurdsson, S. 1993, ApJ, 415, L43

Thorsett, S. E., Arzoumanian, Z., & Taylor, J. H. 1993, ApJ, 412, L33

Wolszczan, A. 1994, talk presented at the Aspen Winter Physics Conf. on Millisecond Pulsars: The Decade of Surprise

Wolszczan, A., & Frail, D. A. 1992, Nature, 355, 145

Fig. 1.— Long-term secular evolution of the binary pulsar's eccentricity for two representative cases. In (a), the second companion is a planet of mass $m_2 = 80\,M_\oplus$ in an orbit of period $P_2 = 10\,\mathrm{yr}$ and eccentricity $e_2 = 0.5$. In (b) it is a star of mass $m_2 = 0.8\,M_\odot$, with $P_2 = 120\,\mathrm{yr}$ and $e_2 = 0.8$. Coplanar orbits have been assumed, and $e_1 = 0$ at $t = 0$. In (a), the eccentricity never grows to a value comparable to what is observed today ($e_1 = 0.025$).

Fig. 2.— Maximum eccentricity induced in the binary pulsar's orbit as a function of the eccentricity of the second companion's orbit. Conventions are as in Fig. 1, except that $e_2$ is varied. The dots are from numerical integrations of the three-body problem, while the solid straight lines show the results of analytic perturbation theory. The vertical dashed lines show the dynamical stability limits. Note that for the planetary case (a), the maximum induced eccentricity remains much smaller than observed today, even when $e_2$ is very close to the limit for stability.

Fig. 3.— Short-term variations of the longitude of pericenter $\omega_1$ and eccentricity $e_1$ for a configuration with $m_2 = 0.8\,M_\odot$, $a_1/r_2 = 10^{-2}$, $\phi_2 = 0.1\pi$, and $\sin\theta_2 = 1$. See text for details. The solid lines show the results of a numerical integration (performed with $m_2$ on a circular orbit). The dashed lines show the analytic results (eqs. [5] and [6], which assume a fixed position for $m_2$).

Fig. 4.— Amplitude of short-term secular variations as a function of the longitude $\phi_2$ of the second companion. All other parameters have the same values as in Fig. 3. The dots are from numerical integrations. The solid lines show the analytic results.